\documentclass{Interspeech}
\interspeechcameraready

\title{ZSDEVC: Zero-Shot Diffusion-based Emotional Voice Conversion with Disentangled Mechanism}

\author[affiliation={1}]{Hsing-Hang}{Chou}
\author[affiliation={1}]{Yun-Shao}{Lin}
\author[affiliation={2}]{Ching-Chin}{Sung}
\author[affiliation={2}]{Yu}{Tsao}
\author[affiliation={1}]{Chi-Chun}{Lee}

\affiliation{Department of Electrical Engineering}{National Tsing Hua University}{Taiwan}
\affiliation{Research Center for Information Technology Innovation}{Academia Sinica}{Taiwan}
\email{stargazer@gapp.nthu.edu.tw, astanley18074@gmail.com, g9612508@gmail.com, yu.tsao@citi.sinica.edu.tw, cclee@ee.nthu.edu.tw}

\usepackage{tabularx}
\usepackage{multicol}
\usepackage{multirow}
\usepackage{amssymb}
\usepackage{amsmath}
\usepackage{svg}
\usepackage{stfloats}

\usepackage{comment}
\usepackage{pifont}
\usepackage{varwidth}
\usepackage{tcolorbox}
\usepackage{pdfpages}

\newcolumntype{Y}{>{\centering\arraybackslash}X}
\let\mc=\multicolumn
\let\mr=\multirow
\let\cl=\cline

\begin{document}

\maketitle

\begin{abstract}
The human voice conveys not just words but also emotional states and individuality. Emotional voice conversion (EVC) modifies emotional expressions while preserving linguistic content and speaker identity, improving applications like human-machine interaction. While deep learning has advanced EVC models for specific target speakers on well-crafted emotional datasets, existing methods often face issues with emotion accuracy and speech distortion. In addition, the zero-shot scenario, in which emotion conversion is applied to unseen speakers, remains underexplored. This work introduces a novel diffusion framework with disentangled mechanisms and expressive guidance, trained on a large emotional speech dataset and evaluated on unseen speakers across in-domain and out-of-domain datasets. Experimental results show that our method produces expressive speech with high emotional accuracy, naturalness, and quality, showcasing its potential for broader EVC applications.
\end{abstract}
\section{Introduction}

Human speech is more than just a medium for words; it conveys an audible declaration of one's identity and emotion \cite{tiwari2012voice}. Modifying emotional expressions while preserving linguistic content and speaker identity is the focus of emotional voice conversion (EVC), a technology that enhances user experiences in applications of human-machine interactions, virtual assistants, and entertainment industries \cite{ZHOU20221, triantafyllopoulos2023overview}. Advances in deep learning are the driving forces in the progression of EVC technology, which has been investigated mainly in the context of speaker-dependent emotion conversion, demonstrating cases of realistically expressive and natural-sounding synthesized speech \cite{zhou20_odyssey, 9054579, zhou21b_interspeech, zhou2022emotion}.

Many deep learning methods have been investigated to achieve emotional voice conversion. These methods fall mainly into two categories: adversarial generative networks (GAN) \cite{zhou20_odyssey, 9054579} and autoencoders \cite{zhou21b_interspeech, zhou2022emotion}. GAN-based approaches use adversarial mechanisms to learn direct mappings between data distributions of different emotional states, allowing direct emotion conversion. In contrast, autoencoder-based methods decompose speech into distinct representation units, such as linguistic content, speaker identity, and emotional information, providing better control over emotion conversion. However, despite their advances, these techniques remain suboptimal in accurately converting emotional states and can introduce distortions in the converted voice, compromising its naturalness and overall quality. The application is further limited due to the data requirements of having well-crafted target speakers' emotional speech samples.

\begin{figure}[t]
    \centering
     \includegraphics[height=2.5cm]{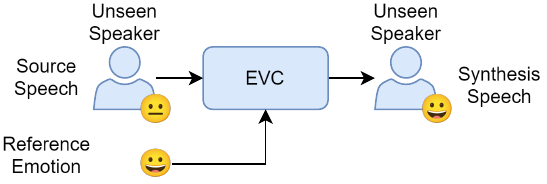}
    \caption{Emotional voice Conversion under the zero-shot scenario, where the emotion state of speech with the unseen speaker is converted.}
  \label{fig:concept}
  \vspace{-1mm}
\end{figure}
A large-scale naturalistic emotional speech corpus such as MSP-Podcast \cite{Lotfian_2019_3} includes a wide range of contexts, speakers, and emotional states. Although originally intended for speech emotion recognition research, recent EVC studies have seen major advancements leveraging these datasets. In particular, Prosody2Vec \cite{qu2023disentangling} achieves high accuracy in converting emotions by learning disentangled prosody representations in various speech datasets, separating emotion information from linguistic content and speaker traits through unsupervised reconstruction. Meanwhile, diffusion models, known for their generative capabilities in producing high-quality samples across multiple applications \cite{song2021scorebased, popovdiffusion, yoon2021adversarial}, have also gained traction in EVC. A prime example is EMOCONV-DIFF \cite{prabhu2024emoconv}, which has significantly improved intensity controllability compared to its predecessor \cite{prabhu2023wild}, while maintaining excellent quality. However, the zero-shot scenario, in which emotion conversion is applied to \textit{ unknown} speakers not present in training data, is underexplored, limiting the generalizability of current models in real-world applications. To fully harness the potential of large-scale emotional speech datasets and enhance the robustness of EVC, further in-depth research into zero-shot scenarios for EVC is crucial.

In contrast to most prior studies that focus on speaker-dependent scenarios, this work aims to develop a zero-shot (unseen speaker) EVC method by introducing a novel diffusion framework with a disentangled mechanism and expressive guidance. The model is trained on the MSP-Podcast dataset, which includes non-parallel real-world emotional speech, and is evaluated on speech from unseen speakers across both in-domain and out-of-domain datasets. To assess the effectiveness of our method, we perform comprehensive objective and subjective evaluations in multiple aspects of synthesized speech, including naturalness, quality, speaker similarity, and emotion classification accuracy, comparing our model with various strong EVC baseline approaches. Our results demonstrate that the proposed zero-shot model performs comparably (often better) to the SOTA EVC models across all metrics, showcasing its promise for broader applications.

\begin{figure*}[ht]
    \centering
    \includegraphics[width=\textwidth,height=5cm]{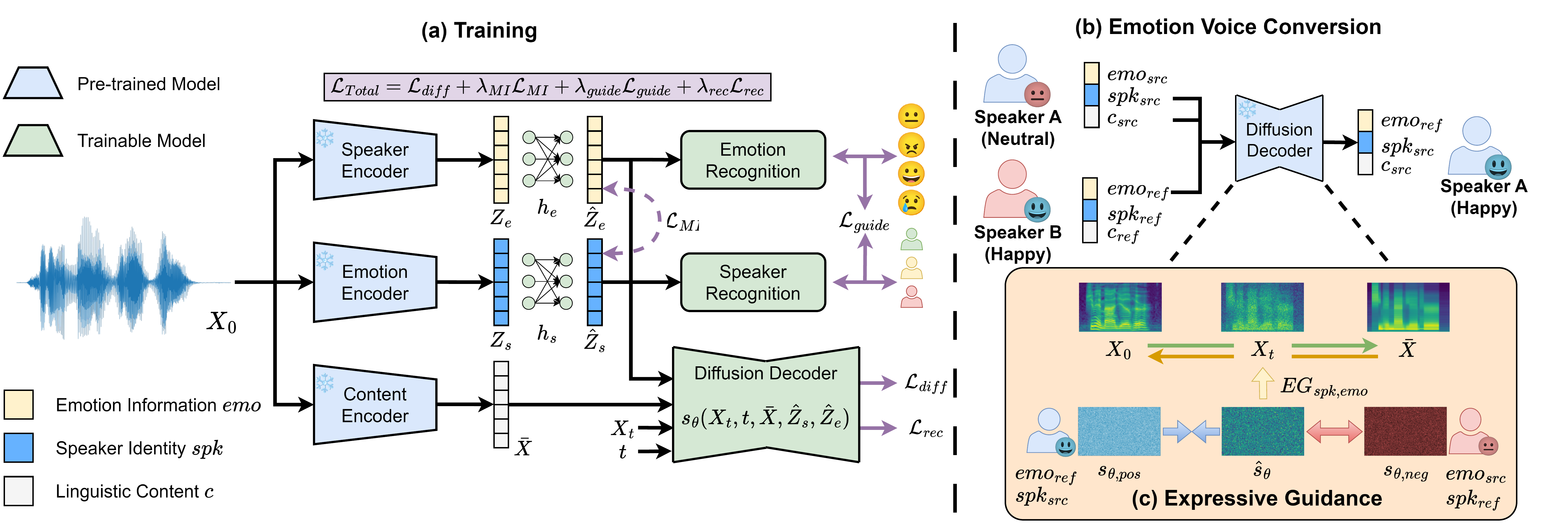}
    \caption{Overview of the proposed diffusion-based zero-shot emotion voice conversion framework.}
  \label{fig:framework}
\end{figure*}

\section{Methodology}
\label{sec:method}
\subsection{Proposed Method}

\label{ssec:architecture}
This work aims to solve the problem of zero-shot emotional voice conversions. Given a pair of $X_{src}:=g(c_{src},spk_{src},emo_{src})$ and reference $X_{ref}:=g(c_{ref},spk_{ref},emo_{ref})$ speech utterances, where each utterance is composed of linguistic content $c$, speaker identity $spk$, and emotion information $emo$ and $g(.)$ is a generative process, our proposed method $G$ aims to perform the conversion process $\hat{X}=G(c_{src}, spk_{src}, emo_{ref})$ that preserves both content and speaker identity while transforming emotion from $emo_{src}$ to $emo_{ref}$. We focus on a zero-shot scenario in which both the source and reference speech, as well as the speaker's identity, remain entirely \emph{unseen} during training.

Figure \ref{fig:framework} illustrates the overall framework of our proposed method. First, multiple encoders extract distinct components with a disentanglement mechanism that ensures their separation. Next, a diffusion-based decoder reconstructs the mel spectrogram based on these components. During inference, a guidance method is applied to push the results from negative to positive conditions. Finally, a pre-trained HiFi-GAN vocoder \cite{kong2020hifi} converts the generated mel spectrogram back to the time-domain signal.

\subsubsection{Encoders}
\label{sssec:encoder}
Three pre-trained encoders are used to capture linguistic content representations $c$, speaker identity $spk$, and emotional expression $emo$.\\
{\bf Phoneme Encoding}: To encode linguistic content $\Bar{X}$, we adapt a pre-trained transformer-based encoder from \cite{popov2021grad} to convert input mel-spectrograms $X_0$ into speaker and emotion independent ``average-voice" mel features that replace each phoneme-level mel feature with the corresponding average phoneme-level mel features.\\
{\bf Speaker Encoding}: To encode the speaker identity $Z_s\in \mathbb{R}^{256}$, we use a pre-trained speaker verification model \cite{jia2018transfer} adapted from \cite{popovdiffusion}.\\
{\bf Emotion Encoding}: To encode emotional information $Z_e\in R^{1024}$, we use an SSL-based SER system adapted from \cite{wagner2023dawn} that was built by fine-tuning the Wav2Vec2-Large-Robutst \cite{baevski2020wav2vec} network on the MSP-Podcast (v1.7) dataset \cite{Lotfian_2019_3}.

To disentangle speaker and emotion representations, we encode the corresponding disentangled representations as $\hat{Z_s}=h_s(Z_s)$ and $\hat{Z_e}=h_e(Z_e)$, where $h_s$ and $h_e$ are linear transformations with learnable parameters.

\subsubsection{Diffusion Decoder}
\label{sssec:diffusion-decoder}
We employ the diffusion framework based on stochastic differential equations (SDE) described in \cite{popovdiffusion}, conditioned on given representations $\bar{X}, Z_s, Z_e$ to generate high-quality speech. The diffusion process gradually transforms the real sample ${X_{0}}$ into ${X_{t}}$ with time-step $t\in[0,1]$ that terminates at average-voice mel-spectrogram $\bar{X}$ when $t=1$ by adding Gaussian noise in a forward process; and generates ${X_{0}}$ from $\bar{X}$ by removing the corresponding score estimation $s_{\theta}(X_t,t,\bar{X},\hat{Z_s},\hat{Z_e})$ in a reverse process. The $s_{\theta}$ with parameter $\theta$ is trained by minimizing mean square error loss $\mathcal{L}_{diff}$ between added noise and $s_{\theta}$.

\subsubsection{Expressive Guidance}
\label{ssec:expressive guidance}
To enhance the effectiveness of the diffusion model on the converted speech, we further design the expressive guidance method that aims to manage the reversed diffusion process with positive and negative direction scores. During the inference stage, we modified $s_\theta$ with $\hat{s}_\theta$ as follows:
\begin{equation} \label{eq:reverse_diff} 
\begin{aligned}
\hat{s}_\theta&=s_{\theta,neg}
+ \lambda_{EG}(s_{\theta,pos}
- s_{\theta,neg})
\end{aligned}
\end{equation}
$\lambda_{EG}$ with the value \textgreater 1 controls the intensity of this guidance method and pushes the generation process away from the negative condition but toward the positive condition. For zero-shot emotion voice conversions, the positive condition takes the source linguistic content $c_{src}$, the source speaker identity $spk_{src}$, and the reference emotion information $emo_{ref}$; On the other hand, the negative condition can be either changing $spk_{src}$ to $spk_{ref}$ for $EG_{spk}$, $emo_{ref}$ to $emo_{src}$ for $EG_{emo}$ or both for $EG_{spk, emo}$, where $EG$ stands for the proposed expressive guidance method.

\subsubsection{Disentangled Loss}
\label{sssec:loss function}

To reduce the correlation between different speech representations, specifically emotion information and speaker identity, We minimize the mutual information (MI) loss between the representations $\mathcal{L}_{MI}=\hat{I}(\hat{z}_s, \hat{z}_e)$, 
where $\hat{I}$ represents the unbiased estimation using vCLUB as described in \cite{cheng2020club}. Prior work has validated the effectiveness of mutual information loss in disentangling speech representations \cite{wang21n_interspeech, yang22f_interspeech}.

To further preserve speaker identity and emotion information residing in the representations after disentanglement, we use two auxiliary supervised models that 1) predict speaker identity from disentangled speaker representation $\hat{z}_s$, and 2) predict emotion labels (Neutral, Angry, Happy, Sad and Surprise) and emotion attributes (Arousal and Valence) from disentangled emotion representation $\hat{z}_e$. These models are trained to minimize loss $\mathcal{L}_{style}$, where the negative log-likelihood loss is used for the categorical prediction task and the concordance correlation coefficient loss is used for the regression task.

In addition to $L_{diff}$ for training diffusion-based decoder, we follow \cite{prabhu2024emoconv} to use a mel-spectrogram recontruction loss $\mathcal{L}_{rec}$ that measures the $\mathcal{L}_{1-norm}$ between $X_0$ and $\hat{X}_0$, where $\hat{X}_0$ is the single-step approximation relying on $X_t, \bar{X}, s_\theta$ using Tweedie's formula \cite{efron2011tweedie}. We use $\lambda_{rec}=(1-t^2)$ adapted from \cite{prabhu2024emoconv} to reduce the importance of the loss as 
$X_t$ becomes increasingly noisy due to added Gaussian noise at larger values of t. 

The final objective function for our proposed method is as follows
\begin{equation} \label{eq:loss_total}
\mathcal{L}_{Total}=\mathcal{L}_{diff}+\lambda_{MI}\mathcal{L}_{MI}+\lambda_{style}\mathcal{L}_{style}+\lambda_{rec}\mathcal{L}_{rec}
\end{equation}
where $\lambda_{MI}$ and $\lambda_{guide}$ are hyparameters to control the importance of respective loss.
\begin{table*}[!t]
\centering
\caption{Objective and subjective evaluation of the proposed method and baseline models on the ESD under either seen or unseen speaker scenarios.}
\resizebox{\textwidth}{!}{
\begin{tabular}{l|c|ccccc|ccc}
\toprule\specialrule{\cmidrulewidth}{0pt}{0pt}
\mr{3}{*}{Method}   &\mr{3}{*}{Scenerio}&\mc{5}{c|}{Objective}&\mc{3}{c}{Subjective}\\
\cl{3-10}
&&\mr{2}{*}{UTMOS}&
\mr{2}{*}{SECS}     &\mc{2}{c}{DNSMOS}&
\mr{2}{*}{ECA}      &\mr{2}{*}{MOS}&
\mr{2}{*}{nMOS}     &\mr{2}{*}{ECA}
\\
\cl{5-6}
&&&&SIG&OVRL&&&&\\
\hline
Target&&
3.606&0.816&3.429&3.155&1.000&
4.197$\pm$0.178&4.402$\pm$0.150&0.889\\ \hline
StarGAN-EVC \cite{9054579}&Seen Speaker&
3.128&\textbf{0.884}&3.461&3.190&0.222& 
4.000$\pm$0.188&3.863$\pm$0.228&0.299\\
Seq2Seq-EVC \cite{zhou21b_interspeech}&Seen Speaker&
1.903&0.663&3.301&2.957&0.444&
1.872$\pm$0.177&2.872$\pm$0.236&0.120\\
Emovox \cite{zhou2022emotion}&Seen Speaker&
2.381&0.698&3.234&2.930&0.333&
2.197$\pm$0.188&2.974$\pm$0.211&0.333\\
Prosody2Vec \cite{qu2023disentangling}&Seen Speaker&
2.482&0.730&3.071&2.717&\textbf{0.889}&
2.803$\pm$0.221&3.308$\pm$0.237&\textbf{0.769}\\ \hline
EMOCONV-DIFF \cite{prabhu2024emoconv}&Unseen Speaker&
\textbf{3.973}&0.834&\textbf{3.611}&\textbf{3.347}&0.667&
\textbf{4.709$\pm$0.098}&\textbf{4.291$\pm$0.179}&0.256\\
ZSDEVC (Proposed)&Unseen Speaker&
\underline{3.583}&\underline{0.768}&\underline{3.589}&\underline{3.336}&\underline{\textbf{0.889}}&
\underline{4.342}$\pm$\underline{0.156}&\underline{3.752}$\pm$\underline{0.201}&\underline{0.530}\\

\specialrule{\cmidrulewidth}{0pt}{0pt}\bottomrule

\end{tabular}
}
\label{table:model_comparison}
\end{table*}

\section{Experimental Setup and Results}
\label{sec:exp}

\subsection{Experimental Setup}
\label{ssec:exp setup}

\subsubsection{Implementation Details}
\label{sssec:implement}
Our proposed methodology is trained on the in-the-wild MSP-Podcast corpus \cite{Lotfian_2019_3} that contains real podcast recordings (16 kHz, 1 ch) with emotional expressions segmented in utterances. We selected 48389 utterances labeled with five emotion labels and emotion attributes from 1381 unique speakers. Each model in ablation studies, including the reimplementation of EMOCONV-DIFF, is trained for 663k iterations with a batch size of 32. The Adam optimizer with a learning rate of $1\times10^{-4}$ is used to update the trainable model parameters. We set $\lambda_{MI}=0.1$ and $\lambda_{style}=1$ during training, and set $\lambda_{EG}=1.25$ for expressive guidance during inference.
\subsubsection{Evaluation Setup}
\label{sssec: eval setup}
We first compared our proposed method with five baseline models: StarGAN-EVC \cite{9054579}, Seq2Seq-EVC \cite{zhou21b_interspeech}, Emovox \cite{zhou2022emotion}, Prosody2Vec \cite{qu2023disentangling}, and EMOCONV-DIFF \cite{prabhu2024emoconv}, using synthesis samples based on act-out emotional speech datatset (ESD) \cite{ZHOU20221} as presented in Prosody2Vec\footnote{https://leyuanqu.github.io/Prosody2Vec/}. Among these, only EMOCONV-DIFF and our method operate in zero-shot scenarios, whereas the other models are either trained or fine-tuned on acted-out ESD. The audio samples are available on our demo page\footnote{https://henrychou36.github.io/ZSDEVC/}.

To assess the effectiveness of our method, we then evaluated our methods in zero-shot scenarios on both in-the-wild datasets, MSP-Podcast, with real-world scenarios, and the act-out dataset, ESD, with high-quality recordings. We randomly sample 300 utterances of each emotion category with unseen speakers from both datasets to conduct the following experiments as source speech. We then perform zero-shot emotional voice conversions that include all the transformations between angry, happy, sad, and neutral, except transforming from emotional speech to neutral. We compared the methods under different training schemes and structures, i.e., using only $\mathcal{L}_{diff}+\mathcal{\lambda}_{rec}\mathcal{L}_{rec}$, which is the reimplementation of EMOCONV-DIFF, and using $\mathcal{L}_{Total}$ in equation \ref{eq:loss_total} with additional layers for disentanglement. We then apply the proposed expressive guidance method on the model trained with $\mathcal{L}_{Total}$. We compared $EG_{spk}$, $EG_{emo}$, and $EG_{spk,emo}$ with different settings of negative condition that replace the representation of positive condition corresponding to speaker identity $spk$, emotion information $emo$, or both, respectively. We also evaluate the Source (VO), which represents the source speech reconstructed by the vocoder.

\subsubsection{Evaluation Metric}
\label{sssec:eval_metric}
For both experiments, we incorporate a non-intrusive objective evaluation, that is, UTMOS \cite{saeki2022utmos} for naturalness, DNSMOS \cite{reddy2022dnsmos} for speech quality (SIG) and overall signal quality (OVRL). Both methods are designed to predict the mean opinion score (MOS) of subjective listening tests. To access speaker similarity, speaker embedding cosine similarity (SECS) between extracted embeddings of source and generated speech based on Resemblyzer \cite{jia2018transfer} is used. For controllability over emotion, we utilized a speech emotion recognition (SER) model fine-tuned on both MSP-Podcast and ESD based on emotion embedding from \cite{wagner2023dawn} to assess emotion classification accuracy (ECA). For the first experiment, in addition to objective evaluation, we conducted a subjective assessment with 13 subjects evaluating 72 converted or target utterances using a 5-point scale ranging from 1 to 5 to assess speech quality and naturalness. We report the mean opinion scores with a 95\% confidence interval for speech quality (MOS) and naturalness (nMOS). The subjects are also required to label the primary emotion for subjective ECA. The evaluation of the first and second experiments is presented in table \ref{table:model_comparison} and table \ref{table:method_comparison} separately. The bolded results indicate the best performance over methodologies, while the underlined results represent our proposed method.

\begin{table*}[t]
\centering
\caption{Objective evaluation of different training and inference schemes of proposed methods for zero-shot emotional voice conversion. We also report the percentage of improvement of ECA compared to the baseline method.}
\resizebox{\textwidth}{!}{%
\begin{tabular}{l|ccccc|ccccc}
\toprule\specialrule{\cmidrulewidth}{0pt}{0pt}
\mr{3}{*}{Methods}&\mc{5}{c|}{MSP-Podcast}&\mc{5}{c}{ESD}\\
\cl{2-11}&\mr{2}{*}{UTMOS}&\mr{2}{*}{SECS}&\mc{2}{c}{DNSMOS}&\mr{2}{*}{ECA}&\mr{2}{*}{UTMOS}&\mr{2}{*}{SECS}&\mc{2}{c}{DNSMOS}&\mr{2}{*}{ECA}\\ 
\cl{4-5}\cl{9-10}&&&SIG&OVRL&&&&SIG&OVRL\\ \hline
Source&
2.830&1.000&3.401&2.892&0.621&
3.927&1.000&3.479&3.191&0.951\\
Source (VO)&
2.488&0.974&3.422&2.907&0.608&
3.538&0.971&3.500&3.210&0.881\\ \hline
EMOCONV-DIFF&
2.477&\textbf{0.837}&\textbf{3.528}&\textbf{3.096}&0.500&
3.708&\textbf{0.822}&\textbf{3.572}&\textbf{3.315}&0.453\\

Our method ($\mathcal{L}_{Total}$)&
2.427&0.773&3.505&3.073&0.584 (16.8$\%\uparrow$)&
3.687&0.763&3.561&3.300&0.548 (21.1$\%\uparrow$)\\
w/ $EG_{spk}$&
\textbf{2.493}&0.788&3.503&3.068&0.557 (11.4$\%\uparrow$)&
\textbf{3.729}&0.774&3.562&3.302&0.504 (11.3$\%\uparrow$)\\
w/ $EG_{emo}$&
2.353&0.744&3.485&3.040&\textbf{0.699} (40.0$\%\uparrow$)&
3.665&0.747&3.560&3.298&\textbf{0.622} (37.4$\%\uparrow$)
\\
w/ $EG_{spk,emo}$&
\underline{2.383}&\underline{0.766}&\underline{3.484}&\underline{3.032}&\underline{0.672} (\underline{34.4}$\%\uparrow$)&
\underline{3.699}&\underline{0.763}&\underline{3.562}&\underline{3.300}&\underline{0.580} (\underline{28.1}$\%\uparrow$)\\
\specialrule{\cmidrulewidth}{0pt}{0pt}\bottomrule
\end{tabular}
}
\label{table:method_comparison}
\end{table*}

\subsection{Experimental Results}

Based on the results in table \ref{table:model_comparison}, our method, as an approach specifically designed for EVC under the zero-shot (\emph{unseen} speaker) scenario, demonstrates overall superior naturalness, quality, and speaker similarity compared to the state-of-the-art EVC method, ProsoV2Vec, in the \emph{seen-speaker} scenario. Additionally, it achieves comparable performance in emotion controllability, as measured by the objective ECA metric, while maintaining distortion-free naturalness similar to the target speech.

By comparing different architectures, we found that autoencoder-based methods such as Prosody2Vec generally achieve higher emotion accuracy than GAN-based methods like StarGAN-EVC. However, they exhibit significantly lower naturalness, quality, and speaker similarity, indicating greater distortion in synthesized samples. On the other hand, diffusion-based methods offer superior naturalness and quality compared to previous approaches, with speaker similarity second only to GAN-based methods, even in \emph{unseen} scenarios. However, their controllability of emotions is less humanly perceptible, as indicated by subjective ECA evaluations. Our method leverages the diffusion model to generate high-quality speech while significantly enhancing emotion controllability in objective and subjective evaluations.

\subsection{Ablation Studies}
Based on table \ref{table:method_comparison}, we observe that simply reconstructing samples using a vocoder introduces artifacts, leading to a degradation in naturalness and reduced emotion recognizability. Compared to EMOCONV-DIFF, which is the backbone diffusion architecture of our methods, our disentanglement mechanism effectively enhances emotion controllability, as reflected in the improved ECA scores—16.8\% for MSP-Podcast and 21.1\% for ESD.

Furthermore, the proposed guidance method significantly improves emotion controllability in terms of ECA during inference, achieving enhancements of 34.4\% for MSP-Podcast and 28.1\% for ESD. Additionally, applying a negative condition to the speaker condition enhances both naturalness and speaker similarity, whereas applying it to the emotion condition substantially boosts emotion accuracy. By focusing on a single condition, the model can be guided to align more precisely with the desired task, whether it be speaker consistency or emotion controllability. Meanwhile, incorporating both conditions ensures balanced performance, maintaining a trade-off between different aspects of speech synthesis quality.
 
Comparing the two datasets in unseen speaker scenarios, we observe that MSP-Podcast, with its real-world samples and complex environments, exhibits lower naturalness and poses greater challenges for emotion recognition, as indicated by UTMOS and ECA, compared to ESD, which consists of high-quality, acted-out recordings. However, our method consistently achieves similar improvements in ECA performance in both datasets, demonstrating its robustness.

\begin{figure}[t]
    \centering
    \includegraphics[width=\columnwidth]{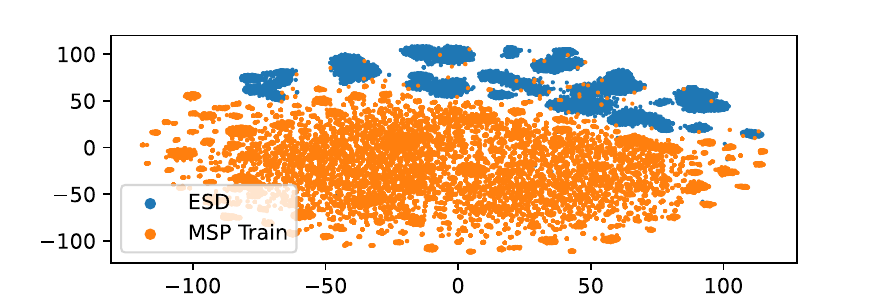}
    \caption{T-SNE plot of speaker embedding over training data of MSP-Podcast and ESD}
  \label{fig:tsne_spk}
  \vspace{-.7cm}
\end{figure}

\subsection{Analysis of Zero-Shot EVC}
To validate how the proposed method enables zero-shot scenario applications, we examined the T-SNE plot of speaker embeddings across the MSP-Podcast samples used for training and the entire ESD, as shown in Figure \ref{fig:tsne_spk}. The two datasets exhibit distinct distributions, confirming that the application is not due to the inclusion of supposedly unseen speakers in the larger dataset. Instead, it likely results from enhanced data diversity in terms of content, speaker characteristics, and emotional states, which facilitates more robust learning and generalization. Moreover, compared to the backbone model, the proposed method further leverages the advantages of the dataset, improving the essential task of emotional voice conversion (EVC) by enhancing emotion controllability.

\section{Conclusion and future work}
\label{sec:conclusion}
In this work, we propose a zero-shot (\emph{unseen speaker}) emotional voice conversion framework that integrates a disentanglement mechanism along with expressive guidance and undergoes a comprehensive objective and subjective evaluation. Our findings reveal several key advantages: (1) the proposed framework effectively enhances accuracy in converting emotion compared to the backbone diffusion-based methods, allowing for more precise modulation of emotional expressions; (2) compared to other EVC frameworks in the \emph{seen speaker scenario}, it produces less distorted emotional speech while maintaining a comparable (often improved) level of emotion controllability; (3) it enables zero-shot scenarios by leveraging the rich diversity of in-the-wild datasets. For future work, we aim to further refine methodologies for zero-shot scenarios to enhance overall performance and robustness.
\bibliographystyle{IEEEtran}
\bibliography{mybib}

\end{document}